\documentclass[review]{elsarticle}
\usepackage{amsfonts}
\usepackage{amsmath}
\usepackage{IEEEtrantools}
\usepackage{lineno,hyperref}
\usepackage{xcolor}
\DeclareMathOperator*{\argmax}{arg\,max}

\newcommand{\ud}{\,\mathrm{d}}
\journal{Physical Communication}

\begin{document}

\begin{frontmatter}

\title{Union Bounds on the Symbol Error Probability of LoRa Modulation for Flat Rician Block Fading Channels}

\author{Rıfat Volkan Şenyuva}
%\address{Marmara Eğitim Köyü, İstanbul}

\author[mymainaddress]{Maltepe University}
%\ead[url]{www.maltepe.edu.tr}

\cortext[mycorrespondingauthor]{Corresponding author}
\ead{rifatvolkansenyuva@maltepe.edu.tr}

\address[mymainaddress]{Marmara Eğitim Köyü, Maltepe, İstanbul}

\begin{abstract}
In this paper the symbol error performance of LoRa modulation is addressed for flat Rician block fading channels. First the exact symbol error probability of the LoRa modulation on Rician fading is derived. Then the upper and lower union bounds are employed on the derived symbol error probability. The proposed bounds are compared against the exact symbol error probability, the numerical evaluation of the symbol error probability and the state-of-art approximation of the LoRa symbol error probability. Numerical results show that while the proposed upper bound is very tight to the exact symbol error probability, there is approximately a 2.5 dB gap for the lower bound.
\end{abstract}


\end{frontmatter}

%\linenumbers

\section{Introduction}
\label{sec:Intro}

The uncoded symbol error performance of the Long Range (LoRa) modulation scheme is investigated for flat Rician block fading channels in this paper. The calculation of the exact symbol error probability of the LoRa modulation is computationally very challenging for the signal dimensions and the constellation sizes used in practice. Instead of direct computation of the symbol error probability, approximations \cite{loraBERPerf,compErrRiceNakagami} have been proposed in the literature to gain an insight into the uncoded symbol error performance under various fading models. We show that the bounds as tight as the approximations but less convoluted can be obtained by employing the lower and upper union bounds for Rician fading.

The Internet of Things (IoT) envisions massive connectivity of devices with low power constraints such as sensors and drones. The communication between such a large number of devices differs from the human-type communication and so is considered as machine-type communication (MTC). The data traffic pattern of the MTC is sporadic, i.e. only a small number of devices transmit short packets at any given time and the transmissions are mostly uplink oriented. Devices transmit without any coordination and there is no resource assignment. Low Power Wide Area Networks (LPWANs) \cite{lpwanPhysical,lpwanSurvey} is a promising solution to these challenges and has been deployed in IoT applications such as environment monitoring \cite{largeAreaMonitoring}, smart irrigation \cite{smartIrrigation}, and unmanned aerial vehicle (UAV) communication systems \cite{UAV_LoRa}. As another key technology of future wireless networks, the UAVs may be used as either a relaying station decoding and forwarding the signals of a ground source reflected by a reconfigurable intelligent surface (RIS) \cite{RIS_dualHop_UAV} or an aerial base station to increase the performance of a dual-hop underwater wireless optical communication (UWOC) system where an autonomous underwater vehicle is receiving transmissions through a over-sea surface relay (R) \cite{UAV_RF-UWOC}. Since the Rician fading model is valid for both the UAV-RIS link in the RIS assisted UAV system \cite{RIS_dualHop_UAV} and the UAV-R link in the dual-hop UWOC system \cite{UAV_RF-UWOC}, the proposed union bounds can also be applied in these scenarios if LoRa modulation is adopted in the communication links to the UAVs.

As one of the proprietary physical layer technologies used in LPWANs, LoRa has recently attracted much research interest due to its advantages over conventional modulation schemes. LoRa is a chirp spread-spectrum based $M$-ary modulation scheme. Each LoRa symbol frequency modulates one of $M$ chirp waveforms each with a different initial instantaneous frequency. The number of possible waveforms, $M$, is determined by the spreading factor (SF). SF is a tuneable parameter which allows LoRa to trade data rate for coverage area, link robustness or energy consumption. SF can be increased to extend the coverage area of LoRa up to 5 km in urban areas and up to 15 km in rural areas \cite{lpwanPhysical}. The data rate, on the other hand, is decreased to as low as 300 bps in order to maximize the coverage area. LoRa waveforms have constant envelope which make LoRa robust against narrow-band interference and Doppler offsets \cite{cubeSat}. The phase of the LoRa waveform is continuous \cite{loraSpecAnal} and so stable local clock generation is not required at the demodulator which simplifies the timing and frequency synchronization and reduces the circuitry costs \cite{lpwanPhysical,loraReceiverImplement}. 

The symbol error analysis of the uncoded LoRa modulation is investigated for the additive white Gaussian noise (AWGN) channels in \cite{loraBERPerf,compErrRiceNakagami,baruffa20} and the fading channels in \cite{loraBERPerf,compErrRiceNakagami,LoRaNakagami}. The closed form expressions of the uncoded symbol error probability for Rayleigh, Rician, and Nakagami fading channel models are given in \cite{LoRaNakagami}. The evaluation of these closed form symbol error probability expressions in high-level programming environments such as MATLAB, Octave, and Python \cite{compErrRiceNakagami} is problematic due to precision errors resulting from the computation of the binomial coefficient terms found in these expressions when $\text{SF}>5$ and requires low-level high precision C libraries \cite{baruffa20}. To mitigate this computational issue, \cite{loraBERPerf} employs the Gaussian approximation of the Rician distribution for high signal-to-noise ratios (SNR) on the output of the LoRa matched filter and proposes an approximation to the uncoded symbol error probability for the AWGN and Rayleigh fading channels. Another approach shown in \cite{compErrRiceNakagami} expands the conditional error probability in the uncoded symbol error probability integral into Taylor series and proposes Marcum function based symbol error probability approximations for the AWGN, Rayleigh, Rician, and Nakagami fading channels. This approach \cite{compErrRiceNakagami} introduces a binomial coefficient and its computation loses precision for high SF values such as $\text{SF}=12$. \cite{baruffa20} employs the union bound on the symbol error probability for the AWGN channel but the tightness of the union bound at low SNR is maintained by using a correction factor.  The symbol and frame error analysis of the LoRa modulation with Hamming coding is investigated in \cite{baruffa20,errorRateIntefere,codedFrameAnalysis}. \cite{baruffa20} applies its union approximation on the bit error probability of the coded LoRa modulation with zero-forcing and phase equalization in a multipath channel. \cite{errorRateIntefere} examines a scenario with an interfering LoRa terminal using the same SF. \cite{codedFrameAnalysis} uses the approximation of \cite{loraBERPerf} to find the codeword error probability of the coded LoRa modulation.

The contributions of the paper can be summarized as follows:
\begin{itemize}
	\item To the best of our knowledge, the formulated lower and upper union bounds on the symbol error probability of the uncoded LoRa modulation using the two region approach are first in the literature.
 	\item The derived union bounds do not rely on approximations like the correction term used in the union bound proposed by \cite{baruffa20}, and are applicable to both Rician and Rayleigh fading channels unlike \cite{baruffa20} which can only be applied for AWGN channels.
	\item The proposed union bounds for the Rician fading do not contain any binomial coefficients and so their computational complexity is less than the Rician approximation of \cite{compErrRiceNakagami} which is a Taylor series expansion with a binomial coefficient.
	\item The proposed bounds for the Rayleigh fading consists of only arithmetic and elementary functions unlike the approximations of either \cite{loraBERPerf} which has $Q$ function calls or the approximation of \cite{compErrRiceNakagami} which is a summation with a binomial coefficient and so their computational complexity is less than those of the approximations.
\end{itemize}

The organization for the rest of this paper is in the following way; Section \ref{sec:LoRaSignal} introduces the discrete-time baseband signal model of the LoRa modulation and the detection rule used in the demodulation of LoRa signals. Then the statistics of the LoRa detector are given and the exact symbol error probability is shown in Section \ref{sec:ExactErr}. Section \ref{sec:Union} gives the detailed derivations of the proposed upper and lower union bound on the symbol error probability. The performance of the proposed bounds are compared against the numerical error rate and the approximations of the error rate published in the literature in Section \ref{sec:NumRes}. Finally, Section \ref{sec:Conclusion} emphasizes the results of the paper and discusses extensions for future work.

\section{LoRa Signal Model and Detection Rule}
\label{sec:LoRaSignal}

LoRa modulation is an $M$-ary orthogonal modulation scheme which uses frequency shifted chirp waveforms for baseband modulation. The LoRa encoder converts SF-length tuples of information bits, $( b_{0},...,b_{\text{SF}-1} )$ into $M$ dimensional baseband signal samples, $\mathbf{a}_{m}=[a_{m}[0],...,a_{m}[M-1]]^{T}$, at every LoRa symbol interval, $T_{\text{symbol}}$. The baseband signal dimension is chosen as $M=2^{\text{SF}}$ with SF taking integer values from 7 up to 12 in practice. Due to being an orthogonal modulation, the spectral efficiency of the LoRa encoder is
\begin{equation}
    \rho=\text{SF}/M
    \label{eq:specEff}
\end{equation}
 bits per one complex dimension or bits per two real dimensions. Each symbol in the LoRa constellation $\mathbf{a}_{m}\in\mathcal{A}=\lbrace \mathbf{a}_{0},...,\mathbf{a}_{M-1} \rbrace$ consists of $M$ samples and only one of the samples is nonzero, i.e.
\begin{equation}
    a_{m}[l]=\left\lbrace
                \begin{array}{ll}
                     \sqrt{E}, & l=m  \\
                     0, & \text{else}.
                \end{array}
            \right.
    \label{eq:LoRaSymbol}
\end{equation}
The energy of one LoRa symbol is shown by $E$ (\ref{eq:LoRaSymbol}) and the index of the nonzero sample depends on the encoded bit tuple, i.e. $m=\sum_{j=0}^{\text{SF}-1} b_{j}2^{j}$. The encoded LoRa samples are applied to the input of the LoRa baseband modulator and the LoRa baseband signal is generated as
\begin{equation}
    x_{m}(t)=\sum_{l=0}^{M-1}a_{m}[l]\phi_{l}(t),\qquad 0\le t\le T_{\text{symbol}}.
    \label{eq:LoRaBasebandSignal}
\end{equation}
In (\ref{eq:LoRaBasebandSignal}) $\lbrace \phi_{l}(t)\rbrace$ are the frequency shifted chirp waveforms \cite{loraSpecAnal}
\begin{equation}
    \phi_{l}(t)=\exp\left\lbrace i2\pi Wt\left\lbrack \frac{l}{M}-\frac{1}{2}+\frac{Wt}{2M}-u\left(t-\frac{M-l}{W}\right) \right\rbrack \right\rbrace,\qquad 0\le t\le T_{\text{symbol}}
    \label{eq:LoRaBasebandWave}
\end{equation}
where $W$ is the bandwidth of the chirp waveform and $u(t)$ is the unit step function. The LoRa baseband signal (\ref{eq:LoRaBasebandSignal}) goes through the passband modulator and then the wireless channel resulting in the received signal at the demodulator
\begin{equation}
    Y(t)=2\Re\left\lbrace\left\lbrack H(t)*x_{m}(t)+N(t) \right\rbrack e^{i2\pi f_{c}t} \right\rbrace, \qquad 0\le t\le T_{\text{symbol}}.
    \label{eq:LoRaReceivedPassbandSig}
\end{equation}
In (\ref{eq:LoRaReceivedPassbandSig}) $H(t)$ represents the baseband equivalent channel process, $N(t)$ is the additive white Gaussian baseband noise process with single-sided power spectral density of $N_{0}$, and the carrier frequency is shown as $f_{c}$. The carrier frequencies used in the LoRa passband modulation are in sub-GHz ISM band, i.e. in Europe 863-870 MHz and in the United States 902-928 MHz band \cite{lpwanPhysical}. The received signal is applied to an Hilbert filter and then converted to baseband by multiplying with the conjugate of the analytical carrier signal. The resulting baseband signal is sampled with $T_{\text{sampling}}=1/W$ yielding the discrete time model of $M=WT_{\text{symbol}}$ complex samples as
\begin{equation}
    Y[k]=Hx_{m}[k]+N[k], \qquad k=0,...,M-1.
    \label{eq:LoRaReceivedBasebandSig}
\end{equation}
Since the coherence time and frequency of the wireless channel is assumed to be greater than the symbol interval and the bandwidth respectively, the baseband channel sample is modeled as a Rician random variable with $H\sim \mathcal{CN}(\mu_{H},\sigma_{H}^{2})$ where $\mu_{H}\in\mathbb{C}$. The transmitted baseband LoRa samples in (\ref{eq:LoRaReceivedBasebandSig}) can be written as
\begin{equation}
    x_{m}[k]=x_{m}(kT_{\text{sampling}})=\sqrt{\frac{E}{M}}(-1)^{k}e^{i\pi k^{2}/M}e^{i2\pi mk/M}=x_{0}[k]e^{i2\pi mk/M}
\end{equation}
in terms of the up-chirp signal, $x_{0,k}$, and an exponential tone at $m$-th discrete frequency \cite{loraSpecAnal}. The noise samples in baseband (\ref{eq:LoRaReceivedBasebandSig}) are circularly symmetric Gaussian random variables that is $N[k]\sim\mathcal{CN}(0,N_{0})$.

Since the squared absolute values of the outputs of the baseband demodulators are sufficient statistics, the detection rule for the discrete-time model (\ref{eq:LoRaReceivedBasebandSig}) is given as
\begin{IEEEeqnarray}{rCl}
    \hat{m}&=& \argmax_{0\le \tilde{m}\le M-1} \left| \sum_{k=0}^{M-1} Y[k]\overline{x}_{\tilde{m}}[k] \right|^{2} \nonumber \\
    &=&\argmax_{0\le \tilde{m}\le M-1} \left| \sum_{k=0}^{M-1} \underbrace{Y[k]\overline{x}_{0}[k]}_{V[k]}e^{-i2\pi \tilde{m}k/M} \right|^{2} \nonumber \\
    &=& \argmax_{0\le \tilde{m}\le M-1} \left| \check{V}[\tilde{m}] \right|^{2}
    \label{eq:detectionRule}
\end{IEEEeqnarray}
where the de-chirped received signal is the multiplication of the received signal with the down-chirp signal, i.e. $V[k]=Y[k]\overline{x}_{0}[k]$, and the discrete Fourier transform (DFT) of the de-chirped signal evaluated at $\tilde{m}$-th discrete frequency is shown as $\check{V}[\tilde{m}]=\sum_{k=0}^{M-1}V[k]e^{-i2\pi \tilde{m}k/M}$ \cite{ssk-lora}. The DFT can be implemented using the fast Fourier transform (FFT) algorithm in $O(M\log M)$ complexity.

\section{Symbol Error Probability}
\label{sec:ExactErr}

An erroneous decoding using the detection rule \eqref{eq:detectionRule} occurs when $m$ is the transmitted symbol and the squared absolute value of the DFT of de-chirped signal has its maximum value at a frequency other than $m$. If $R_{\tilde{m}}=|\check{V}[\tilde{m}]|^{2}$ is defined, then the probability density function (PDF), $f_{R_{m}}(r_{m})$, of $R_{m}$ can be used to find the probability of a symbol error conditional on $m$ being transmitted as
\begin{equation}
    \textrm{Pr}(\varepsilon|\mathbf{A}=\mathbf{a}_{m})=\textrm{Pr}(\varepsilon)=\int_{0}^{\infty} f_{R_m}(r_m)\textrm{Pr}\left\lbrack\bigcup_{\tilde{m}\neq m}(R_{\tilde{m}}>r_m)\right\rbrack\ud r_m.
    \label{eq:symbolErr}
\end{equation}
 Due to the symmetry of the constellation, the conditional symbol error \eqref{eq:symbolErr} is the same for all symbols in the constellation and so is equal to the unconditional symbol error probability. In order to evaluate \eqref{eq:symbolErr}, the distribution of $R_{m}$ has to be determined and so the distribution of $\check{V}[\tilde{m}]$ has to be found first.

Conditional on symbol $m$ being transmitted, the DFT of the de-chirped signal \eqref{eq:detectionRule} can be rewritten as
\begin{IEEEeqnarray}{rCl}
    \check{V}[\tilde{m}]&=& \sum_{k=0}^{M-1}Y_{m}[k]\overline{x}_{0}[k]e^{-i2\pi \tilde{m}k/M} \nonumber\\
    &=&\sum_{k=0}^{M-1}(Hx_{0}[k]e^{i2\pi mk/M}+N[k])\overline{x}_{0}[k]e^{-i2\pi \tilde{m}k/M} \nonumber \\
    &=&\frac{E}{M}H\sum_{k=0}^{M-1}e^{i2\pi (m-\tilde{m})k/M}+\sum_{k=0}^{M-1}N[k]\overline{x}_{0}[k]e^{-i2\pi \tilde{m}k/M} \nonumber \\
    &=&\left\lbrace
        \begin{array}{ll}
             EH+\tilde{N}[\tilde{m}], & \tilde{m}=m  \\
             \tilde{N}[\tilde{m}], & \tilde{m}\neq m
        \end{array}\right.
    \label{eq:decisionMetricForMessage}
\end{IEEEeqnarray}
where $\tilde{N}[\tilde{m}]$ is circularly symmetric Gaussian random variable with $\tilde{N}[\tilde{m}]\sim\mathcal{CN}(0,EN_{0})$ and $H$ is the Rician random variable representing the single-tap channel. Thus, the distribution of $\check{V}[\tilde{m}]$ is given as
\begin{equation}
    \check{V}[\tilde{m}]=\left\lbrace
        \begin{array}{ll}
            \mathcal{CN}(E\mu_{H},E^{2}\sigma_{H}^{2}+EN_{0}), & \tilde{m}=m \\
            \mathcal{CN}(0,EN_{0}), & \tilde{m}\neq m.
        \end{array}\right.
    \label{eq:metricDist}
\end{equation}
As it can be observed from \eqref{eq:metricDist}, for $\tilde{m}\neq m$, $|\check{V}[\tilde{m}]|^{2}$ or $R_{\tilde{m}}$ is the sum of two independent and identically distributed zero mean Gaussian random variables with common variance $EN_{0}/2$ that is
\begin{equation}
    R_{\tilde{m}}=|\check{V}[\tilde{m}]|^{2}=|\check{V}_{\text{re}}[\tilde{m}]+i\check{V}_{\text{im}}[\tilde{m}]|^{2}=|\check{V}_{\text{re}}[\tilde{m}]|^{2}+|\check{V}_{\text{im}}[\tilde{m}]|^{2}
    \label{eq:chiSquare}
\end{equation}
where both the real and imaginary parts of $\check{V}[\tilde{m}]$ are distributed as $\check{V}_{\text{re}}[\tilde{m}]$, $\check{V}_{\text{im}}[\tilde{m}]\sim N(0,EN_{0}/2)$. $|\check{V}[\tilde{m}]|^{2}$ \eqref{eq:chiSquare} is the Chi-square random variable with 2 degrees of freedom and its distribution is equal to that of an exponential random variable with mean $EN_{0}$ \cite{proakisDigital,gallagerComm}. When $\tilde{m}=m$, $|\check{V}[m]|^{2}$ becomes a noncentral Chi-square random variables of 2 degrees of freedom with $s=E|\mu_{H}|$ and common variance of $\sigma_{0}^{2}=(E^{2}\sigma_{H}^{2}+EN_{0})/2$ \cite{proakisDigital,gallagerComm}.

To calculate the symbol error probability given in \eqref{eq:symbolErr}, $\textrm{Pr}[\cup_{\tilde{m}\neq m}(R_{\tilde{m}}>r_m)]$ has to be calculated. We can subtract the probability of the complement event that is $\cap_{\tilde{m}\neq m} R_{\tilde{m}}<r_{m}$ from 1. Since every $R_{\tilde{m}}$ conditioned on $\tilde{m}\neq m$ is exponentially distributed with identical mean $EN_{0}$, the probability of each $R_{\tilde{m}}<r_{m}$ can be found using the exponential PDF \cite{gallagerComm} as
\begin{IEEEeqnarray}{rCl}
    \textrm{Pr}[R_{\tilde{m}}<r_{m}]&=&\int_{0}^{r_{m}} \frac{1}{EN_{0}}\exp\left(-\frac{r'_{m}}{EN_{0}}\right)\ud r'_{m}=\frac{-EN_{0}}{EN_{0}}\left\lbrack \exp\left(\frac{-r_{m}}{EN_{0}}\right)-1\right\rbrack \nonumber \\
                                    &=&1-\exp\left( -\frac{r_{m}}{EN_{0}} \right).
    \label{eq:expoProb}
\end{IEEEeqnarray}
Every $R_{\tilde{m}}$ is independent and so $\textrm{Pr}[\bigcap_{\tilde{m}\neq m}(R_{\tilde{m}}<r_m)]$ is equal to the $(M-1)$-th power of \eqref{eq:expoProb} that is
\begin{equation}
    \textrm{Pr}\left[\bigcap_{\tilde{m}\neq m}(R_{\tilde{m}}<r_m)\right]=\left[1-\exp\left( -\frac{r_{m}}{EN_{0}} \right)\right]^{M-1}.
    \label{eq:complementProb}
\end{equation}
Using \eqref{eq:complementProb}, $\textrm{Pr}[\cup_{\tilde{m}\neq m}(R_{\tilde{m}}>r_m)]$ can be rewritten as 
\begin{equation}
    \textrm{Pr}\left[\bigcup_{\tilde{m}\neq m}(R_{\tilde{m}}>r_m)\right]=1-\left[1-\exp\left( -\frac{r_{m}}{EN_{0}} \right)\right]^{M-1}.
    \label{eq:condErrorProb}
\end{equation}
When \eqref{eq:condErrorProb} is plugged into \eqref{eq:symbolErr}, the symbol error probability can be found from
\begin{equation}
    \textrm{Pr}(\varepsilon)=\int_{0}^{\infty} \left\lbrace 1-\left[1-\exp\left( -\frac{r_{m}}{EN_{0}} \right)\right]^{M-1}\right\rbrace f_{R_m}(r_m)\ud r_m
    \label{eq:numericSymbolErr}
\end{equation}
where the PDF of the noncentral Chi-square random variable of 2 degrees of freedom, $R_{m}$, is given in terms of the modified Bessel function of the first kind and order zero, $I_{0}(\cdot)$, as
\begin{equation}
    f_{R_{m}}(r_{m})=\frac{1}{2\sigma_{0}^2}\exp\left(-\frac{s^{2}+r_{m}}{2\sigma_{0}^{2}} \right)I_{0}\left( \frac{s\sqrt{r_{m}}}{\sigma^{2}_{0}} \right).
\end{equation}
\eqref{eq:numericSymbolErr} cannot be solved analytically but if the binomial expansion \cite{proakisDigital} is used to expand \eqref{eq:complementProb} as
\begin{equation}
    \left[1-\exp\left( -\frac{r_{m}}{EN_{0}} \right)\right]^{M-1}=\sum_{n=0}^{M-1}(-1)^{n}\binom{M-1}{n}\exp\left(\frac{-nr_{m}}{EN_{0}} \right),
    \label{eq:binomExp}
\end{equation}
then the symbol error probability \eqref{eq:symbolErr} can be evaluated from
\begin{IEEEeqnarray}{rCl}
    \textrm{Pr}(\varepsilon)&=&\int_{0}^{\infty} f_{R_m}(r_m)\left\lbrack 1-\sum_{n=0}^{M-1}(-1)^{n}\binom{M-1}{n}\exp\left(\frac{-nr_{m}}{EN_{0}} \right) \right\rbrack\ud r_m \nonumber \\
    &=& 1-\sum_{n=0}^{M-1}(-1)^{n}\binom{M-1}{n}\int_{0}^{\infty}f_{R_{m}}(r_{m})\exp\left(\frac{-nr_{m}}{EN_{0}} \right)\ud r_m \nonumber \\
    &=& \sum_{n=1}^{M-1}(-1)^{n+1}\binom{M-1}{n}\int_{0}^{\infty}f_{R_{m}}(r_{m})\exp\left(\frac{-nr_{m}}{2\sigma_{1}^{2}} \right)\ud r_m
    \label{eq:symbolErrII}
\end{IEEEeqnarray}
where $\sigma_{1}^{2}=EN_{0}/2$. The integral in \eqref{eq:symbolErrII} can be solved in rectangular coordinates. If the real and imaginary parts of $\check{V}[\tilde{m}]$ are shown as $\check{V}_{\text{re}}$ and $\check{V}_{\text{im}}$ respectively, the joint PDF of $\check{V}_{\text{re}}$ and $\check{V}_{\text{im}}$ is given as
\begin{equation}
    f_{\check{V}_{\text{re}},\check{V}_{\text{im}}}({\check{v}_{\text{re}}},{\check{v}_{\text{im}}})=\frac{1}{2\pi \sigma_{0}^{2}}\exp\left\lbrack-\frac{(\check{v}_{\text{re}}-\mu_{1})^{2}}{2\sigma_{0}^{2}}-\frac{(\check{v}_{\text{im}}-\mu_{2})^{2}}{2\sigma_{0}^{2}}\right\rbrack.
    \label{eq:jointRealImag}
\end{equation}
While the means of the real and imaginary parts in \eqref{eq:jointRealImag} are $\mu_{1}=E\Re\lbrace \mu_{H}\rbrace$ and $\mu_{2}=E\Im\lbrace \mu_{H} \rbrace$ respectively, the variances are equal to $\sigma_{0}^{2}$. Using \eqref{eq:jointRealImag}, the integral in \eqref{eq:symbolErrII} is equal to the following integral
\begin{equation}
    \iint\frac{1}{2\pi \sigma_{0}^{2}}\exp\left\lbrack-\frac{(\check{v}_{\text{re}}-\mu_{1})^{2}}{2\sigma_{0}^{2}}-\frac{(\check{v}_{\text{im}}-\mu_{2})^{2}}{2\sigma_{0}^{2}}\right\rbrack \exp\left\lbrack\frac{-n(\check{v}^{2}_{\text{re}}+\check{v}^{2}_{\text{im}})}{2\sigma_{1}^{2}} \right\rbrack\ud\check{v}_{\text{re}}\ud\check{v}_{\text{im}}
    \label{eq:integRectCord}
\end{equation}
We have two Gaussian integrals in \eqref{eq:integRectCord} and they can be solved separately by using the axiom that the area underneath any PDF is equal to one. The integral \eqref{eq:integRectCord} is solved as in 
\begin{IEEEeqnarray}{rCl}
    &\int& \exp\left\lbrack -\frac{(\check{v}_{\text{im}}-\mu_{2})^{2}}{2\sigma_{0}^{2}}-\frac{n\check{v}^{2}_{\text{im}}}{2\sigma_{1}^{2}}  \right\rbrack \ud\check{v}_{\text{im}}  \int\frac{1}{2\pi \sigma_{0}^{2}}\exp\left\lbrack-\frac{(\check{v}_{\text{re}}-\mu_{1})^{2}}{2\sigma_{0}^{2}}-\frac{n\check{v}^{2}_{\text{re}}}{2\sigma_{1}^{2}} \right\rbrack\ud\check{v}_{\text{re}} \nonumber \\
    &=& \frac{\sqrt{2\pi\sigma_{0}^{2}\sigma_{1}^{2}}}{2\pi\sigma_{0}^{2}\sqrt{\sigma_{1}^{2}+n\sigma_{0}^{2}}}\exp\left\lbrack \frac{-n\mu_{1}^{2}}{2(\sigma_{1}^{2}+n\sigma_{0}^{2})} \right\rbrack \int \exp\left\lbrack -\frac{(\check{v}_{\text{im}}-\mu_{2})^{2}}{2\sigma_{0}^{2}}-\frac{n\check{v}^{2}_{\text{im}}}{2\sigma_{1}^{2}}  \right\rbrack \ud\check{v}_{\text{im}} \nonumber \\
    &=&\frac{\sigma_{1}^{2}}{\sigma_{1}^{2}+n\sigma_{0}^{2}}\exp\left\lbrack \frac{-n(\mu_{1}^{2}+\mu_{2}^{2})}{2(\sigma_{1}^{2}+n\sigma_{0}^{2})} \right\rbrack \nonumber \\
    &=&\frac{1}{(n+1)+n\sigma_{H}^{2}(E/N_{0})}\exp\left( -\frac{n(E/N_{0})|\mu_{H}|^{2}}{(n+1)+n\sigma_{H}^{2}(E/N_{0})} \right).
    \label{eq:resIntegRectCord}
\end{IEEEeqnarray}
via completing the squares. Plugging \eqref{eq:resIntegRectCord} in \eqref{eq:symbolErrII}, the symbol error probability can be rewritten to get the error probability in \cite{LoRaNakagami} that is
\begin{equation}
    \textrm{Pr}(\varepsilon)=\sum_{n=1}^{M-1} \frac{(-1)^{n+1}}{(n+1)+n\sigma_{H}^{2}(E/N_{0})}\binom{M-1}{n}\exp\left( -\frac{n(E/N_{0})|\mu_{H}|^{2}}{(n+1)+n\sigma_{H}^{2}(E/N_{0})} \right).
    \label{eq:symbolErrIII}
\end{equation}
If the mean and variance of the channel tap are taken as $\mu_{H}=1$ and $\sigma_{H}^{2}=0$ respectively, the error probability \eqref{eq:symbolErrIII} reduces to the noncoherent case given in \cite{proakisDigital,loraBERPerf}
\begin{equation}
    \textrm{Pr}(\varepsilon)=\sum_{n=1}^{M-1} \frac{(-1)^{n+1}}{n+1}\binom{M-1}{n}\exp\left( -\frac{n}{n+1}\frac{E}{N_{0}} \right).
    \label{eq:symbolErrNoncoherent}
\end{equation}
where the magnitude of the channel tap, $|H|$, is assumed to be unit constant ($\sigma_{H}^{2}=0$) while its phase $\angle H$ is assumed to be uniformly distributed between $(0,2\pi)$. As $M$ gets larger, the computation of the binomial coefficient in \eqref{eq:symbolErrIII} becomes more challenging. This led to the approximate expressions for \eqref{eq:symbolErrIII} such as the one derived in \cite{loraBERPerf} for the Rayleigh fading. Section \ref{sec:Union} shows how union bounds can be employed to get tight bounds on \eqref{eq:symbolErrIII}.

\section{Union Bounds on Symbol Error Probability}
\label{sec:Union}

This section investigates the application of union bounds for the symbol error probability derived in \eqref{eq:symbolErrIII}. Instead of direct calculation of the probability of the union of the error events in \eqref{eq:symbolErr}, the union bound can be used to bound this probability as
\begin{equation}
    (M-1)p-\frac{(M-1)^{2}p^{2}}{2}\le \textrm{Pr}\left\lbrack\bigcup_{\tilde{m}\neq m}(R_{\tilde{m}}>r_m)\right\rbrack \le (M-1)p.
    \label{eq:unionBound}
\end{equation}
where $p=\textrm{Pr}[R_{\tilde{m}}>r_{m}]$ and using \eqref{eq:expoProb} $p$ can be found as
\begin{equation}
    p=1-\textrm{Pr}[R_{\tilde{m}}<r_{m}]=\exp\left( -\frac{r_{m}}{EN_{0}} \right).
    \label{eq:errEventProb}
\end{equation}
If \eqref{eq:errEventProb} is plugged in \eqref{eq:unionBound}, the upper union bound can be rewritten as 
\begin{equation}
    \textrm{Pr}\left\lbrack\bigcup_{\tilde{m}\neq m}(R_{\tilde{m}}>r_m)\right\rbrack \le (M-1)\exp\left( -\frac{r_{m}}{EN_{0}} \right).
    \label{eq:unionUpper}
\end{equation}
Since the exponential function is monotonic decreasing in its argument, the right hand side of the inequality \eqref{eq:unionUpper} is going to be much larger than the obvious bound of 1 for any probability for small $r_{m}$ . Using this fact, \eqref{eq:unionUpper} can be made tighter as in
\begin{equation}
    \textrm{Pr}\left\lbrack\bigcup_{\tilde{m}\neq m}(R_{\tilde{m}}>r_m)\right\rbrack \le \left\lbrace
    \begin{array}{ll}
       (M-1)\exp( -r_{m}/EN_{0}),  & r_{m}>r^{*}_{m} \\
       1, & r_{m}\le r^{*}_{m}
    \end{array}
    \right.
    \label{eq:unionUpRobust}
\end{equation}
where $r^{*}_{m}$ is given as
\begin{equation}
    r^{*}_{m}=EN_{0}\ln(M-1)
    \label{eq:integralLimit}
\end{equation}
By plugging \eqref{eq:unionUpRobust} in \eqref{eq:symbolErr}, the upper bound on the symbol error probability can be evaluated as
\begin{equation}
    \text{Pr}(\varepsilon)\le\int_{0}^{r^{*}_{m}}f_{R_{m}}(r_{m})\ud r_{m}+(M-1)\int_{r^{*}_{m}}^{\infty}f_{R_{m}}(r_{m})\exp\left(-\frac{r_{m}}{EN_{0}}\right)\ud r_{m}
    \label{eq:upperBound1}
\end{equation}
The first integral in \eqref{eq:upperBound1} is equal to the cumulative distribution function of the noncentral Chi-square random variable with 2 degrees of freedom, $F_{R_{m}}(r^{*}_{m})$, which is given as
\begin{equation}
    \int_{0}^{r^{*}_{m}}f_{R_{m}}(r_{m})\ud r_{m}=F_{R_{m}}(r^{*}_{m})=1-Q_{1}\left( \frac{s}{\sigma_{0}},\frac{\sqrt{r^{*}_{m}}}{\sigma_{0}} \right)
    \label{eq:upBoundInt1}
\end{equation}
where $Q_{1}$ is the Marcum $Q$ function \cite{proakisDigital}. The second integral in \eqref{eq:upperBound1} can be written explicitly
\begin{IEEEeqnarray}{rCl}
    &\int_{r^{*}_{m}}^{\infty}&\frac{1}{2\sigma_{0}^{2}}I_{0}\left(\frac{s}{\sigma_{0}^{2}}\sqrt{r_{m}}\right)\exp\left(-\frac{s^{2}+r_{m}}{2\sigma_{0}^{2}}\right)\exp\left(-\frac{r_{m}}{EN_{0}}\right)\ud r_{m}= \nonumber \\
    &\int_{r^{*}_{m}}^{\infty}&\frac{1}{2\sigma_{0}^{2}}I_{0}\left(\frac{s}{\sigma_{0}^{2}}\sqrt{r_{m}}\right)\exp\left(-\frac{s^{2}+r_{m}}{2\sigma_{0}^{2}}-\frac{r_{m}}{2\sigma_{1}^{2}}\right)\ud r_{m}= \nonumber \\
    &\int_{r^{*}_{m}}^{\infty}&\frac{1}{2\sigma_{0}^{2}}I_{0}\left(\frac{s}{\sigma_{0}^{2}}\sqrt{r_{m}}\right)\exp\left(-\frac{s^{2}+\left(1+\frac{\sigma_{0}^{2}}{\sigma_{1}^{2}}\right)r_{m}}{2\sigma_{0}^{2}} \right)\ud r_{m}= \nonumber \\
    e^{-\frac{s^{2}}{2(\sigma_{0}^{2}+\sigma_{1}^{2})}}&\int_{r^{*}_{m}}^{\infty}&\frac{1}{2\sigma_{0}^{2}}I_{0}\left(\frac{s}{\sigma_{0}^{2}}\sqrt{r_{m}}\right)\exp\left(-\frac{\frac{s^{2}}{\left(1+\frac{\sigma_{0}^{2}}{\sigma_{1}^{2}}\right)}+(1+\frac{\sigma_{0}^{2}}{\sigma_{1}^{2}})r_{m}}{2\sigma_{0}^{2}} \right)\ud r_{m}= \nonumber \\
    \frac{e^{-\frac{s^{2}}{2(\sigma_{0}^{2}+\sigma_{1}^{2})}}}{1+\sigma_{0}^{2}/\sigma_{1}^{2}}&\int_{\tilde{r}^{*}}^{\infty}&\frac{1}{2\sigma_{0}^{2}}I_{0}\left(\frac{\tilde{s}}{\sigma_{0}^{2}}\sqrt{\tilde{r}}\right)\exp\left(-\frac{\tilde{s}^{2}+\tilde{r}}{2\sigma_{0}^{2}} \right)\ud \tilde{r}= \nonumber \\
    \frac{e^{-\frac{s^{2}}{2(\sigma_{0}^{2}+\sigma_{1}^{2})}}}{1+\sigma_{0}^{2}/\sigma_{1}^{2}}&Q_{1}&\left( \frac{\tilde{s}}{\sigma_{0}},\frac{\sqrt{\tilde{r}^{*}}}{\sigma_{0}} \right)
    \label{eq:upBoundInt2}
\end{IEEEeqnarray}
where
\begin{equation}
    \tilde{s}=\frac{s}{1+\sigma_{0}^{2}/\sigma_{1}^{2}}, \qquad \tilde{r}^{*}=(1+\sigma_{0}^{2}/\sigma_{1}^{2})r^{*}_{m}.
\end{equation}
Plugging \eqref{eq:upBoundInt1} and \eqref{eq:upBoundInt2} into \eqref{eq:upperBound1} yields the upper bound on the symbol error probability as
\begin{IEEEeqnarray}{rCl}
    \text{Pr}(\varepsilon)&\le&1-Q_{1}\left( \frac{s}{\sigma_{0}},\frac{\sqrt{r^{*}_{m}}}{\sigma_{0}} \right)+\frac{(M-1)}{1+\sigma_{0}^{2}/\sigma_{1}^{2}}e^{-\frac{s^{2}}{2(\sigma_{0}^{2}+\sigma_{1}^{2})}}Q_{1}\left( \frac{\tilde{s}}{\sigma_{0}},\frac{\sqrt{\tilde{r}^{*}}}{\sigma_{0}} \right) \nonumber \\
    &\le& 1-Q_{1}(\alpha_{1},\beta_{1})+\frac{M-1}{2+\sigma_{H}^{2}(E/N_{0})}e^{-\frac{|\mu_{H}|^{2}}{\sigma_{H}^{2}+2(E/N_{0})^{-1}}}Q_{1}(\alpha_{2},\beta_{2})
    \label{eq:upperBound2}
\end{IEEEeqnarray}
where the arguments of the first Marcum $Q$ function are
\begin{IEEEeqnarray}{rCl}
    \alpha_{1}&=&\frac{s}{\sigma_{0}}=\sqrt{\frac{2|\mu_{H}|^{2}}{\sigma_{H}^{2}+(E/N_{0})^{-1}}} \label{eq:FirstMarcumArg1} \\ \beta_{1}&=&\frac{\sqrt{r^{*}_{m}}}{\sigma_{0}}=\sqrt{\frac{2\ln{(M-1)}}{1+\sigma_{H}^{2}(E/N_{0})}}
    \label{eq:FirstMarcumArg2}
\end{IEEEeqnarray}
and the arguments of the second Marcum $Q$ function are
\begin{IEEEeqnarray}{rCl}
    \alpha_{2}&=&\frac{\tilde{s}}{\sigma_{0}}=\sqrt{\frac{2|\mu_{H}|^{2}}{3\sigma_{H}^{2}+2(E/N_{0})^{-1}+(\sigma_{H}^{2})^{2}(E/N_{0})}} \label{eq:SecMarcumArg1} \\
    \beta_{2}&=&\frac{\sqrt{\tilde{r}^{*}}}{\sigma_{0}}=\sqrt{2\ln{(M-1)}\left\lbrack 1+\frac{1}{1+\sigma_{H}^{2}(E/N_{0})} \right\rbrack}.
    \label{eq:SecMarcumArg2}
\end{IEEEeqnarray}
For a given $E/N_{0}$, the complexity of computing \eqref{eq:upperBound2} is 54 floating-point operations (FLOPs) plus twice the computational complexity of the Marcum Q function. A FLOP is considered as one addition, subtraction, multiplication, division of two floating-point numbers. The complexity of taking the square root, the exponential, and the logarithm of a floating number is also equal to 1 FLOP. Finding the square of the absolute value of a complex number is assumed to be 3 FLOPs. Using the Marcum $Q$ function bound, i.e. $Q_{1}(\alpha_{2},\beta_{2})\le e^{-\frac{(\beta_{2}-\alpha_{2})^{2}}{2}}$, from \cite{simonFading}, the upper union bound \eqref{eq:upperBound2} can be further bounded from above 
\begin{equation}
     \text{Pr}(\varepsilon)\le 1-Q_{1}(\alpha_{1},\beta_{1})+\frac{M-1}{2+\sigma_{H}^{2}(E/N_{0})}\exp\left\lbrack-\frac{|\mu_{H}|^{2}}{\sigma_{H}^{2}+2(E/N_{0})^{-1}}-\frac{(\beta_{2}-\alpha_{2})^{2}}{2}\right\rbrack
     \label{eq:upperBound3}
\end{equation}
for $\beta_{2}>\alpha_{2}\ge 0$.

For $\mu_{H}=0$, the Rician random variable reduces to a Rayleigh random variable and the upper union bound \eqref{eq:upperBound2} can be further simplified. When $\mu_{H}=0$, the first arguments of both of the Marcum $Q$ functions \eqref{eq:FirstMarcumArg1},\eqref{eq:SecMarcumArg1} becomes zero, i.e. $\alpha_{1}=\alpha_{2}=0$ and so the Marcum $Q$ functions reduce to exponentials \cite{simonFading} as
\begin{IEEEeqnarray}{rCl}
    Q_{1}(\alpha_{1}=0,\beta_{1})&=&\exp\left(-\frac{\beta_{1}^{2}}{2} \right)=\exp\left\lbrack -\frac{\ln(M-1)}{1+\sigma_{H}^{2}(E/N_{0})} \right\rbrack \label{eq:redMarcum1Up} \\
    Q_{1}(\alpha_{2}=0,\beta_{2})&=&\exp\left(-\frac{\beta_{2}^{2}}{2} \right)=\exp\left\lbrack -\ln(M-1)-\frac{\ln(M-1)}{1+\sigma_{H}^{2}(E/N_{0})} \right\rbrack.
    \label{eq:redMarcum2Up}
\end{IEEEeqnarray}
If \eqref{eq:redMarcum1Up} and \eqref{eq:redMarcum2Up} is plugged into \eqref{eq:upperBound2}, the upper union bound for the Rayleigh case ($\mu_{H}=0$) is obtained as
\begin{equation} 
    \text{Pr}(\varepsilon)\le 1+\left\lbrack \frac{1}{2+\sigma_{H}^{2}(E/N_{0})}-1 \right\rbrack\exp\left\lbrack-\frac{\ln(M-1)}{1+\sigma_{H}^{2}(E/N_{0})} \right\rbrack.
    \label{eq:upRayleighBound}
\end{equation}
The complexity of the upper union bound for the Rayleigh case \eqref{eq:upRayleighBound} is equal to 13 FLOPs.

For the lower bound, the left side of the union bound \eqref{eq:unionBound} is used
\begin{equation}
    \textrm{Pr}\left\lbrack\bigcup_{\tilde{m}\neq m}(R_{\tilde{m}}>r_m)\right\rbrack \ge (M-1)e^{ -\frac{r_{m}}{EN_{0}}}-\frac{(M-1)^{2}}{2}e^{ -\frac{2r_{m}}{EN_{0}}}
    \label{eq:unionLow}
\end{equation}
If $z(r_{m})=(M-1)e^{-r_{m}/EN_{0}}$ is defined, \eqref{eq:unionLow} can be rewritten as
\begin{equation}
    \textrm{Pr}\left\lbrack\bigcup_{\tilde{m}\neq m}(R_{\tilde{m}}>r_m)\right\rbrack \ge z(r_{m})-z^{2}(r_{m})/2.
    \label{eq:unionLow2}
\end{equation}
$z(r_{m})$ \eqref{eq:unionLow2} is decreasing in $r_{m}$ and becomes unit for $r_{m}=r^{*}_{m}$ \eqref{eq:integralLimit}. Using $r^{*}_{m}$ \cite{gallagerComm}, \eqref{eq:unionLow2} can be split into
\begin{equation}
    \textrm{Pr}\left\lbrack\bigcup_{\tilde{m}\neq m}(R_{\tilde{m}}>r_m)\right\rbrack \ge 
                \left\lbrace \begin{array}{ll}
                              \frac{z(r_{m})}{2}=\frac{M-1}{2}e^{-r_{m}/EN_{0}}, & r_{m}>r^{*}_{m} \\
                              1/2,      &  r_{m}\le r^{*}_{m}.
                             \end{array}\right.
    \label{eq:unionLow3}
\end{equation}
For $r_{m}>r^{*}_{m}$, $z^{2}<z<1$ and so $z-z^{2}/2>z/2$ which is the upper part of \eqref{eq:unionLow3}. When $r_{m}=r^{*}_{m}$, $z-z^{2}/2=1/2$ and this result is also valid for $r_{m}<r^{*}_{m}$ since the probability we're trying to bound is increasing with decreasing $r_{m}$. By plugging \eqref{eq:unionLow3} in \eqref{eq:symbolErr}, the lower bound on the symbol error probability can be obtained from
\begin{equation}
    \text{Pr}(\varepsilon)\ge \frac{1}{2}\int_{0}^{r^{*}_{m}}f_{R_{m}}(r_{m})\ud r_{m}+\frac{M-1}{2}\int_{r^{*}_{m}}^{\infty}f_{R_{m}}(r_{m})\exp\left(-\frac{r_{m}}{EN_{0}}\right)\ud r_{m}.
    \label{eq:lowerBound1}
\end{equation}
We' ve already solved the integrals in \eqref{eq:lowerBound1} during the derivation of the upper bound. Plugging \eqref{eq:upBoundInt1} and \eqref{eq:upBoundInt2} into \eqref{eq:lowerBound1} yields
\begin{equation}
    \text{Pr}(\varepsilon)\ge
    \frac{1}{2}[1-Q_{1}(\alpha_{1},\beta_{1})]+\frac{M-1}{4+2\sigma_{H}^{2}(E/N_{0})}e^{-\frac{|\mu_{H}|^{2}}{\sigma_{H}^{2}+2(E/N_{0})^{-1}}}Q_{1}(\alpha_{2},\beta_{2})
    \label{eq:lowerBound2}
\end{equation}
where the arguments of the Marcum functions are the same as in \eqref{eq:FirstMarcumArg1}, \eqref{eq:FirstMarcumArg2}, \eqref{eq:SecMarcumArg1}, \eqref{eq:SecMarcumArg2}. The complexity of the union lower bound \eqref{eq:lowerBound2} is 56 FLOPs plus twice the computational complexity of the Marcum Q function. If the lower bound for the Marcum $Q$ function \cite{simonFading}, i.e. $Q_{1}(\alpha_{2},\beta_{2})\ge e^{-\frac{(\beta_{2}+\alpha_{2})^{2}}{2}}$, is employed, the lower union bound \eqref{eq:lowerBound2} can be rewritten as
\begin{IEEEeqnarray}{rCl}
     \text{Pr}(\varepsilon)&\ge& \frac{1-Q_{1}(\alpha_{1},\beta_{1})}{2} \nonumber \\
     &+&\frac{M-1}{4+2\sigma_{H}^{2}(E/N_{0})}\exp\left\lbrack-\frac{|\mu_{H}|^{2}}{\sigma_{H}^{2}+2(E/N_{0})^{-1}}-\frac{(\beta_{2}+\alpha_{2})^{2}}{2}\right\rbrack
     \label{eq:lowerBound3}
\end{IEEEeqnarray}
for $\beta_{2}>\alpha_{2}\ge 0$.

When $\mu_{H}=0$ is chosen, the Rician variable becomes a Rayleigh variable and so the lower bound expression of \eqref{eq:lowerBound2} can be further simplified. \eqref{eq:redMarcum1Up} and \eqref{eq:redMarcum2Up} can be plugged into \eqref{eq:lowerBound2} yielding the lower union bound for the Rayleigh case as
\begin{equation}
     \text{Pr}(\varepsilon) \ge \frac{1}{2}+\frac{1}{2}\left\lbrack \frac{1}{2+\sigma_{H}^{2}(E/N_{0})}-1 \right\rbrack\exp\left\lbrack-\frac{\ln(M-1)}{1+\sigma_{H}^{2}(E/N_{0})} \right\rbrack.
    \label{eq:lowRayleighBound}
\end{equation}
The complexity of the lower union bound for the Rayleigh case is 14 FLOPs.

\section{Numerical Results}
\label{sec:NumRes}

The spectral efficiency of the LoRa modulation \eqref{eq:specEff} is less than 2 bits per two real dimensions due to being an orthogonal signalling scheme. Thus, we are in the power-limited regime and so the performances should be given in bit error probabilities, $P_{b}(\varepsilon)$, versus the ratio of signal energy per bit to the noise energy per two real dimensions, $E_{b}/N_{0}$. All of the error probabilities derived in sections \ref{sec:ExactErr} and \ref{sec:Union} are symbol error probabilities. The symbol error probabilities can be converted into bit error probabilities using the fact that approximately half of the bits will be erroneous in the event of a symbol error \cite{proakisDigital} as
\begin{equation}
    P_{b}(\varepsilon)=\frac{2^{\text{SF}-1}}{2^{\text{SF}}-1}\text{Pr}(\varepsilon)\approx \frac{1}{2}\text{Pr}(\varepsilon).
\end{equation}
As for the $E_{b}/N_{0}$ ratio, all symbol error probability expressions have $E/N_{0}$ in them. Since $E$ is the energy of a LoRa symbol and a LoRa symbol consists of SF number of bits, the symbol error energy can be written in terms of the bit energy as $E=\text{SF} E_{b}$ and this relationship can be used in the error probability expressions to represent $E/N_{0}=\text{SF}(E_{b}/N_{0})$. To maintain equality between the powers of the transmitted and received signals, the channel gain of the single-tap Rician channel random variable is normalized that is $E\lbrace |H|^{2} \rbrace=\sigma_{H}^{2}+|\mu_{H}|^{2}=1$. The investigated $E_{b}/N_{0}$ range is between [0,40] dB and since $\beta_{2}>\alpha_{2}\ge 0$ holds within this range, the bounds \eqref{eq:upperBound3}, \eqref{eq:lowerBound3} are also applicable. All the bit error probabilities are calculated in a MATLAB environment of R2022a release with version number 9.12.0.2039608 and 5th update installed on a PC. The OS of the PC is 64-bit Windows 11 Pro with AMD Ryzen 5 3600 6-Core Processor at 3.60 GHz and 16 GB RAM. The detector results obtained using the detection rule \eqref{eq:detectionRule} are averaged over $10^{8}$ LoRa symbols.

\begin{figure}[htbp]
 	\centering
 	\includegraphics[width=0.95\textwidth]{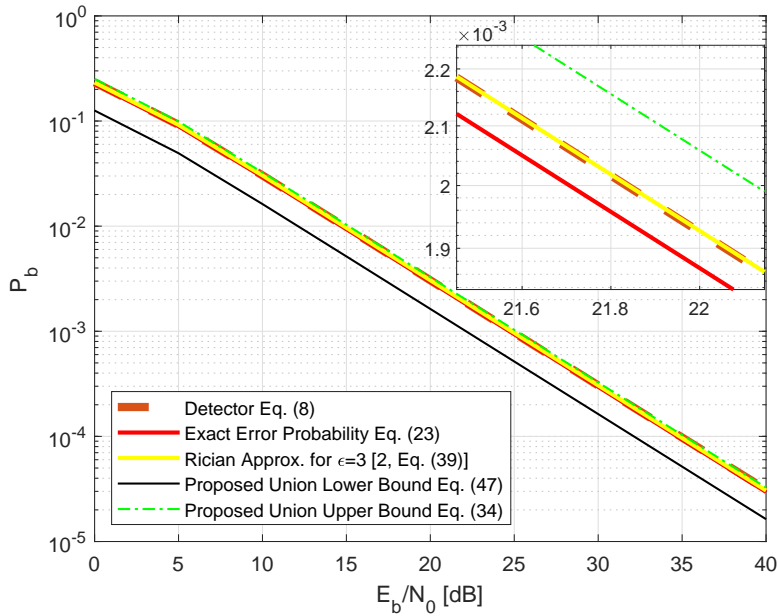}
 	\caption{LoRa bit error performance on single-tap Rician channel for SF=5 ($M=32$).}
 	\label{fig:RicianSF5}
\end{figure}

For the first numerical result given in Figure \ref{fig:RicianSF5}, $\text{SF}$ is chosen as $\text{SF}=5$ equating the baseband signal dimension to $M=32$ so that the binomial coefficient in the exact probability of error \eqref{eq:symbolErrIII} can be evaluated without any precision errors. The mean and the variance of the single-tap Rician channel are chosen as $|\mu_{H}|=\sqrt{1/2}$ and $\sigma_{H}^{2}=1/2$ respectively which results in unit Rician factor or shaping parameter, i.e. $K=|\mu_{H}|^{2}/\sigma_{H}^{2}=1$. As it is observed from Figure \ref{fig:RicianSF5}, the detector result \eqref{eq:detectionRule}, the exact error probability \eqref{eq:symbolErrIII}, and the Rician approximation of \cite{compErrRiceNakagami} using Taylor series order of $\epsilon=3$ are very close to each other. While the union upper bound \eqref{eq:upperBound2} tracks the exact error probability closely, the union lower bound \eqref{eq:lowerBound2} has approximately 2.5 dB gap for the considered $E_{b}/N_{0}$ range.

\begin{figure}[htbp]
 	\centering
 	\includegraphics[width=\textwidth]{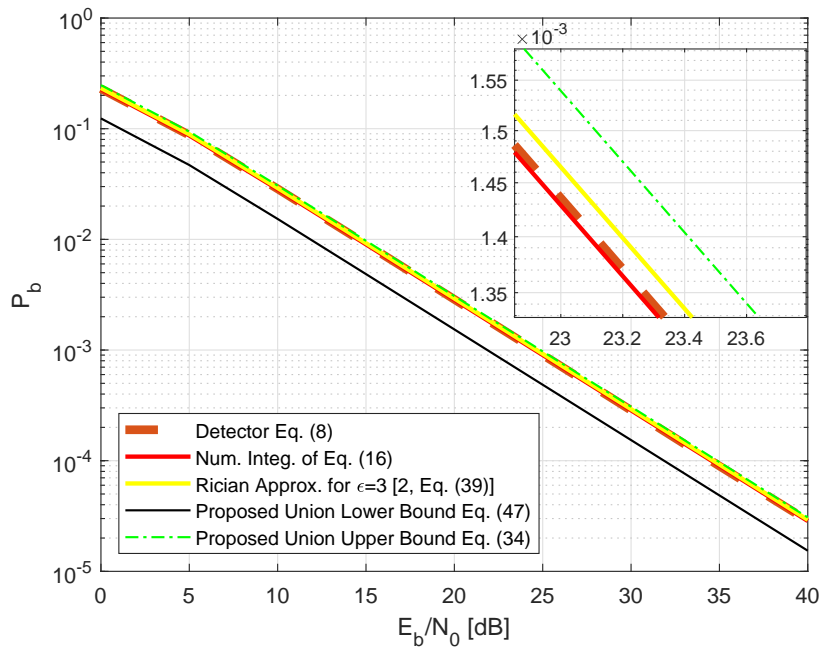}
 	\caption{LoRa bit error performance on single-tap Rician channel for SF=7 ($M=128$).}
 	\label{fig:RicianSF7}
\end{figure}

For the second numerical result (Figure \ref{fig:RicianSF7}), the Rician channel parameters are kept the same, i.e. $|\mu_{H}|=\sqrt{1/2}$ and $\sigma_{H}^{2}=1/2$, but the SF is increased to $\text{SF}=7$ yielding $M=128$ which prevents the usage the exact error probability \eqref{eq:symbolErrIII} due to precision errors in the computation of the binomial coefficient. The numerical integration of the symbol error probability \eqref{eq:numericSymbolErr} is used in place of the exact probability \eqref{eq:symbolErrIII} computation for simulation setups with $\text{SF}>5$. As it can be seen from Figure \ref{fig:RicianSF7}, while the union upper bound \eqref{eq:upperBound2} and the Rician approximation of \cite{compErrRiceNakagami} are still very tight to the numerical integration of the error probability \eqref{eq:numericSymbolErr}, the gap between the union lower bound \eqref{eq:lowerBound2} and the numerical integration remains unchanged across the examined $E_{b}/N_{0}$ range.

\begin{figure}[htbp]
 	\centering
 	\includegraphics[width=\textwidth]{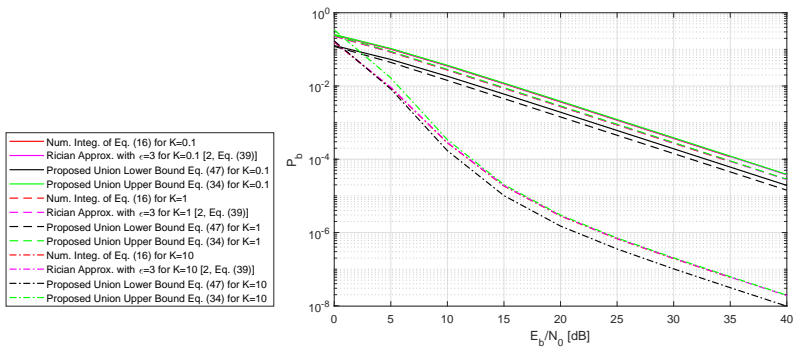}
 	\caption{LoRa bit error performance on single-tap Rician channel with varying $K\in\lbrace 0.1,1,10 \rbrace$ for SF=12 ($M=4096$).}
 	\label{fig:RicianSF12}
\end{figure}

The Rician channel model consists of two terms with a first term corresponding to a line of sight path often called a specular path and a second term corresponding to the aggregation of a large number of reflected and scattered paths. The ratio of the energy in the specular path to the energy in the scattered paths is determined by the shaping parameter or the Rician factor, $K$, of the Rician model. Figure \ref{fig:RicianSF12} compares the proposed union bounds \eqref{eq:lowerBound2},\eqref{eq:upperBound2} against the numerical integration of \eqref{eq:numericSymbolErr} and the Rician approximation of \cite{compErrRiceNakagami} with Taylor series of order $\epsilon=3$ for varying $K\in\lbrace 0.1,1,10 \rbrace$. As $K$ gets smaller, i.e. $K=0.1$, the energy of the specular path gets smaller and the Rician model reduces to a Rayleigh model. For $K\ge 10$, the energy of the line of sight path becomes dominant and the Rician random variable can now be well approximated by a Gaussian and the Rician channel behaves like an AWGN channel.

\begin{figure}[htbp]
 	\centering
 	\includegraphics[width=0.95\textwidth]{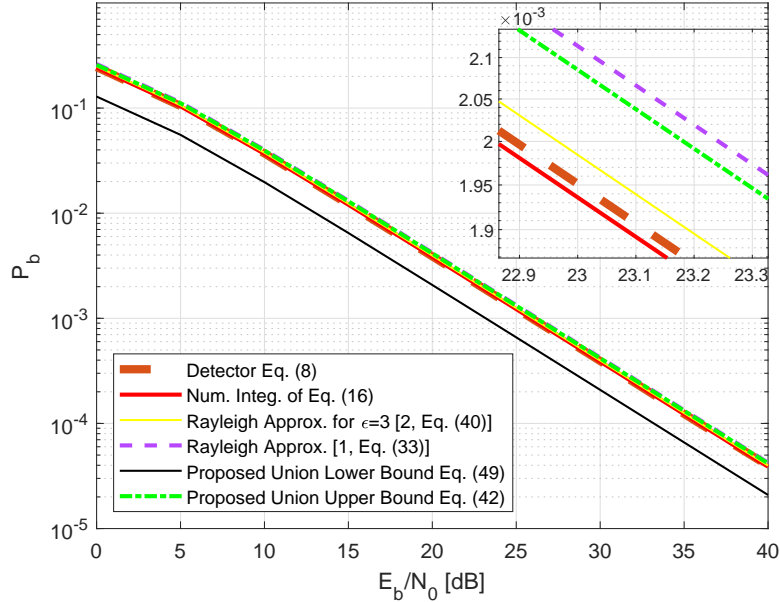}
 	\caption{LoRa bit error performance on single-tap Rayleigh channel for SF=7 ($M=128$).}
 	\label{fig:rayleighSF7}
\end{figure}

The upper \eqref{eq:upRayleighBound} and lower bounds \eqref{eq:lowRayleighBound} derived for the Rayleigh case are compared against the numerical integration of the symbol error probability \eqref{eq:numericSymbolErr} along with the approximations of \cite{loraBERPerf} and \cite{compErrRiceNakagami} in Figure \ref{fig:rayleighSF7}. For the flat Rayleigh block fading channel simulations, the mean is set to zero, i.e. $\mu_{H}=0$, and the variance of the channel is set to unity, i.e. $\sigma_{H}^{2}=1$. Figure \ref{fig:rayleighSF7} shows that while the proposed upper union bound \eqref{eq:upRayleighBound} is as tight as the approximation of \cite{loraBERPerf} to the numerical integration but the results of the Rayleigh approximation of \cite{compErrRiceNakagami} is tighter.

\section{Conclusion}
\label{sec:Conclusion}
This paper examines the application of the union bounds to derive less convoluted computable tight bounds on the uncoded symbol error probability of the LoRa modulation on flat Rician and flat Rayleigh block fading channels. As a future work the proposed bounds on the symbol error probability can be extended to the analysis of frame errors with Hamming coding. Another very important extension problem is the error performance at a LoRa gateway when an interfering LoRa terminal transmits a packet at the same time.

\bibliography{mybibfile}

\end{document}